\documentclass{article}

\usepackage{arxiv}

\usepackage[hidelinks]{hyperref}       
\usepackage{makecell}
\usepackage{amsmath}
\usepackage{graphicx}
\usepackage[super]{natbib}
\bibpunct{}{}{,}{s}{}{\textsuperscript{,}}

\title{SciOps: Achieving Productivity and Reliability in Data-Intensive Research}

\author{
  {\bf Erik C. Johnson}$^{1,a}$ \and
  {\bf Thinh T.\ Nguyen}$^{2}$ \and
  {\bf Benjamin K.\ Dichter}$^{3}$ \and
  {\bf Frank Zappulla}$^{4}$ \and
  {\bf Montgomery Kosma}$^{2}$ \and
  {\bf Kabilar Gunalan}$^{2,5}$ \and
  {\bf Yaroslav O.\ Halchenko}$^{6}$ \and
  {\bf Shay Q.\ Neufeld}$^{7}$ \and
  {\bf Kristen Ratan}$^{8}$ \and
  {\bf Nicholas J.\ Edwards}$^{9}$ \and
  {\bf Susanne Ressl}$^{10}$ \and
  {\bf Sarah R.\ Heilbronner}$^{11}$ \and
  {\bf Michael Schirner}$^{\text{12-16}}$ \and
  {\bf Petra Ritter}$^{\text{12-16}}$ \and
  {\bf Brock Wester}$^{1}$ \and
  {\bf Satrajit Ghosh}$^{5, 16}$ \and
  {\bf Maryann E.\ Martone}$^{17}$ \and
  {\bf Franco Pestilli}$^{18}$ \and
  {\bf Dimitri Yatsenko}$^{2,b}$ 
\\ 
  $^{1}$ Research and Exploratory Development Department, \\
  Johns Hopkins University Applied Physics Laboratory, Laurel, MD, USA \\
  $^{2}$ DataJoint Inc., Houston, TX, USA \\
  $^{3}$ CatalystNeuro, Benicia, CA, USA \\
  $^{4}$ Digital R\&D Creation Center, Pfizer Inc., USA \\
  $^{5}$ McGovern Institute for Brain Research, Massachusetts Institute of Technology, Cambridge, MA, USA \\
  $^{6}$ Center for Open Neuroscience, Department of Psychological and Brain Sciences, \\
  Dartmouth College, New Hampshire, USA \\
  $^{7}$ Inscopix, a Bruker company, Mountain View, CA, USA \\
  $^{8}$ Strategies for Open Science (Stratos), Santa Cruz, CA, USA \\
  $^{9}$ Happy Potato, Inc., Issaquah, WA, USA \\
  $^{10}$ Department of Neuroscience, University of Texas at Austin, Austin, TX \\
  $^{11}$ Neurosurgery, Baylor College of Medicine, Houston, TX \\
  $^{12}$ Berlin Institute of Health (BIH) at Charité – Universitätsmedizin Berlin, Berlin, Germany \\
  $^{13}$ Department of Neurology with Experimental Neurology, Charité, \\ Universitätsmedizin Berlin, Corporate Member of Freie Universität Berlin \\
  and Humboldt Universität zu Berlin, Berlin, Germany \\
  $^{14}$ Bernstein Focus State Dependencies of Learning and \\
  Bernstein Center for Computational Neuroscience, Berlin, Germany \\
  $^{15}$ Einstein Center for Neuroscience Berlin, Berlin, Germany \\
  $^{16}$ Einstein Center Digital Future, Berlin, Germany \\
  $^{16}$ Department of Otolaryngology, Harvard Medical School, Boston, MA, USA \\
  $^{17}$ Department of Neurosciences, University of California San Diego, La Jolla, CA, USA \\
  $^{18}$ Department of Psychology, University of Texas Austin, Austin, TX, USA \\
  $^a$\href{mailto:erik.c.johnson@jhuapl.edu}{\tt erik.c.johnson@jhupl.edu}; $^b$\href{mailto:dimitri@datajoint.com}{\tt dimitri@datajoint.com}
}

\renewcommand{\shorttitle}{SciOps: Achieving Productivity and Reliability}

\hypersetup{
pdftitle={SciOps: Achieving Productivity and Reliability in Data-Intensive Research}
pdfsubject={q-bio.NC, q-bio.QM},
pdfauthor={Erik C.\ Johnson, Thinh T.\ Nguyen, Benjamin K.\ Dichter, Frank Zappulla, Montgomery Kosma, Kabilar Gunalan, Yaroslav O.\ Halchenko, Shay Q.\ Neufeld, Kristen Ratan, Susanne Ressl, Sarah R.\ Heilbronner, Michael Schirner, Petra Ritter, Brock Wster, Satrajit Ghosh, Maryann E.\ Martone, Franco Pestilli, Dimitri Yatsenko,}
pdfkeywords={SciOps, DevOps, DataOps, MLOps, Capability Maturity Model, neuroscience, operations research, open science, closed-loop experiments, digital twin, FAIR, reproducible research, automated workflows, artificial intelligence, AI-driven discovery},
}

\begin{document}
\maketitle

\begin{abstract}
Scientists are increasingly leveraging advances in instruments, automation, and collaborative tools to scale up their experiments and research goals, leading to new bursts of discovery.
Various scientific disciplines, including neuroscience, have adopted key technologies to enhance collaboration, reproducibility, and automation.
Drawing inspiration from advancements in the software industry, we present a roadmap to enhance the reliability and scalability of scientific operations for diverse research teams tackling large and complex projects.
We introduce a five-level Capability Maturity Model describing the principles of rigorous scientific operations in projects ranging from small-scale exploratory studies to large-scale, multi-disciplinary research endeavors.
Achieving higher levels of operational maturity necessitates the adoption of new, technology-enabled methodologies, which we refer to as ``SciOps.''
This concept is derived from the DevOps methodologies that have revolutionized the software industry.
SciOps involves digital research environments that seamlessly integrate computational, automation, and AI-driven efforts throughout the research cycle—from experimental design and data collection to analysis and dissemination, ultimately leading to closed-loop discovery.
This maturity model offers a framework for assessing and improving operational practices in multidisciplinary research teams, guiding them towards greater efficiency and effectiveness in scientific inquiry.
\end{abstract}


\section{The Need for Elevating Operational Maturity in Large-Scale Science}
Research thrives on creativity, exploration, and freedom.
In moments unbound by strict methodologies---such as setting up a lab, starting a Ph.D.\ project, or initiating new experiments---scientists embrace improvisation.
They modify protocols, learn through iteration, and refine techniques and gain insights through hands-on practice and direct observation of unfiltered measurements.
Such experimentation, often disparaged as mere ``tinkering,'' has been the bedrock of scientific inquiry.
History offers numerous examples of intuition-led creativity sparking breakthroughs, including Nobel Prize-winning achievements \cite{norrby2016nobel}.

However, the initial phase of discovery, characterized by exuberant productivity and minimal discipline, often faces significant challenges in credibility and scalability.
Custom, specialized methods may prove successful within their originating laboratory but fail to generalize across wider applications \cite{baker2016reproducibility}.
The need to reproduce and integrate procedures and methods across labs highlights the requirement for balancing innovation with structured, reliable processes.
Scalability---the ability to manage and expand increasingly complex projects predictably---requires robust collaboration, advanced technology, rigorous quality control, and stringent standards for transparency and reproducibility.
In other words, it demands \emph{operational maturity}.

Combining scientific experiments with advanced data science demands broad multidisciplinary collaboration, where teams often struggle to integrate unique technical and organizational methods.
The shift from individual, exploratory research to large-scale, collaborative projects drives the need for teams to demonstrate high operational maturity, maintaining rigorous standards while continuing to innovate.
Enhancing collaboration, streamlining processes, reducing errors, and emphasizing computational and experimental reproducibility are crucial for this transition.

Neuroscience in particular---our primary area of research---is undergoing a transformative shift toward large-scale, data-centric collaborative efforts, driven by advanced neurotechnologies and experimental techniques.
This transformation is rooted in the need for a comprehensive approach that integrates structural and functional organization of neural circuits with molecular interactions and ethological aspects.
Exemplifying this shift are landmark projects such as the Human Connectome Project, which compiles thousands of human brain connectivity maps \cite{van2013wu}, the MICrONS program's detailed mapping of cortical circuits \cite{microns2021functional}, and the BRAIN Initiative Cell Consensus Network's (BICCN) first comprehensive reference of cell types in mouse with the BRAIN Initiative Cell Atlas Network's (BICAN) extending this to humans and non-human primates \cite{brain2021multimodal}.
These projects have not only advanced data acquisition and analysis but also necessitated the development of community standard and coordination strategies.

This scaling of neuroscience has been propelled by significant funding from the US BRAIN Initiative \cite{insel2013nih}, the EU Human Brain Project \cite{amunts2016human}, and the China Brain Project \cite{poo2016china}, philanthropic, and other initiatives.
As datasets and teams grow across various modalities and domains \cite{brinkmann2009large,lichtman2014big,landhuis2017neuroscience}, the potential for groundbreaking discoveries in neuroscience grows---but it cannot be fully realized without a commensurate improvement in effective and scalable scientific operations.

Smaller labs, led by individual principal investigators---still the bedrock of fundamental research---also face the challenge of scaling their activities in data collection and analysis.
Despite smaller teams and budgets, these labs must harness high-bandwidth recording and stimulation techniques, design multimodal experiments, and apply advanced analysis methods.
They can enhance their capabilities by making effective use of open-source projects, open standards, scientific platforms, and community resources, while maintaining greater creative freedom.

As large-scale research becomes increasingly quantitative and digital, scientific manuscripts often remain disconnected from the research process itself.
Researchers must shift their focus from data and analysis to writing, which can create a gap that hinders real-time data sharing, reuse, and reproducibility. However, there is a clear opportunity to develop tools that integrate research communication directly into the workflow, streamlining the process and enhancing collaboration, transparency, and efficiency.

Neuroscience faces distinct challenges in operational maturity compared to fields such as bioinformatics or astrophysics.
The discipline relies on diverse data modalities and instruments across different species and scales, with many core acquisition technologies still rapidly evolving with significant investment from large research initiatives.
These factors pose significant hurdles to standardization and community coordination.
Additionally, neuroscience experiments span a wide range of spatial and temporal scales, involving various species, brain regions, and behaviors, further complicating data operations and reproducibility.
Historically, operational maturity in neuroscience has lagged behind bioinformatics and genomics, where initiatives such as the Human Genome Project set early precedents for scalable data operations.

A new turning point is the accelerating integration of \emph{artificial intelligence} (AI) into research practices \cite{wang2023scientific, national2022automated}.
AI's ability to manage and interpret large datasets is poised to significantly advance the study of complex, dynamic, and adaptive systems such as the brain.
This integration depends on first establishing formal processes in scientific workflows, blending human creativity with AI advancements to enable closed-loop scientific discovery.

This paper establishes operational principles that ensure scalability and integrity in data-intensive research, providing a roadmap to help research teams achieve ambitious goals. 
Drawing inspiration from transformative practices in engineering, business, and software development, we present strategies to enhance operational maturity and support the advancement of increasingly complex scientific endeavors.

\section{Successful Transformations in Operational Maturity}
We examine transformative ideas that have combined human ingenuity with systematic processes to dramatically increase productivity in the software industry.
Our goal is to apply these innovations to neuroscience, where enhanced operations are urgently needed, and to other data-rich, computation-intensive scientific fields.

The first key concept is DevOps, a methodology that has transformed software development over the past 15 years by combining development and operations into a continuous, semi-automated workflow \cite{ebert2016devops, leite2019survey, teixeira2020maturity}.
This approach allows updates, new features, and improvements to be deployed rapidly, often several times a day, without disrupting services.
DevOps uses technologies such as containerization, version control, and infrastructure as code (IaC) to create processes that 1) test small units of functionality, 2) continuously integrate updates to ensure system integrity, and 3) automatically deploy changes to production systems.

These tools and processes automate repetitive tasks, enhance team collaboration, reduce errors, and accelerate project timelines.
DevOps has played a critical role in the rise of the Software-as-a-Service (SaaS) industry, enabling dynamic services such as Netflix, Zoom, Spotify, Slack, and Atlassian.
Importantly, modern platforms and frameworks have democratized DevOps, making its powerful methodologies accessible to teams of all sizes, including small startups, enabling them to implement mature operations efficiently.

Building on DevOps, new methodologies---\textbf{DataOps} and \textbf{MLOps}---have emerged to improve data analytics and machine learning \cite{gartner, rodriguez2020good, atwal2019practical, makinen2021needs}.
These practices enhance teamwork and deliver more accurate insights by reducing delays caused by handoffs between teams.
Instead, they encourage direct collaboration through a unified workflow that supports continuous improvement.
With standardized testing in place, teams can work simultaneously on a semi-automated pipeline, ensuring updates and new developments are smoothly integrated and delivered without interruptions.
In both DataOps and MLOps, new machine learning models are tested and deployed efficiently, incorporating fresh data while maintaining seamless service.

The time has now come for \textbf{SciOps}, a transformative approach to scientific operations that promises to bring similar revolutionary impacts to experimental science as seen in other "Ops" disciplines. 
While the term ``SciOps'' has been used in various contexts previously, we define it in alignment with other ``-Ops'' methodologies, emphasizing streamlined, technology-driven collaboration.
SciOps integrates computation, lab automation, and AI across the entire research cycle---from experiment design to data analysis---accelerating the journey from inquiry to insight.
This approach encompasses all aspects of research, including designing experimental conditions, data collection, simulation, modeling, and exploratory analysis, enabling continuous experimentation cycles.
The 2022 National Academies consensus study on {\em Automated Research Workflows} (a term closely related to SciOps) highlights the adoption of SciOps in various data-intensive fields \cite{national2022automated}.

The second key concept, \textbf{Capability Maturity Model Integration} (CMMI), provides a framework for assessing and enhancing the operational maturity of software development teams.
Developed by the Software Engineering Institute at Carnegie Mellon University, CMMI categorizes teams into five levels---from {\em Initial} to {\em Optimizing}---based on their effectiveness in executing complex projects \cite{paulk2009history,chrissis2011cmmi}.
This model serves as a diagnostic tool, aiding in strategic planning and the evaluation of teams for significant projects.

To conduct high-throughput studies and enhance the reliability of results, research teams need a systematic and incremental approach to improving operational maturity.
Our objective is to integrate principles from Capability Maturity Model Integration (CMMI) and DevOps into research workflows.
This framework aims to scale operations, foster collaboration, and accelerate the pace of scientific discovery, particularly by integrating artificial intelligence with human-driven activities.

\section{The Capability Maturity Model for Science Operations}

\begin{figure*}[htb]
	\centering
	\includegraphics[width=\textwidth=180mm]{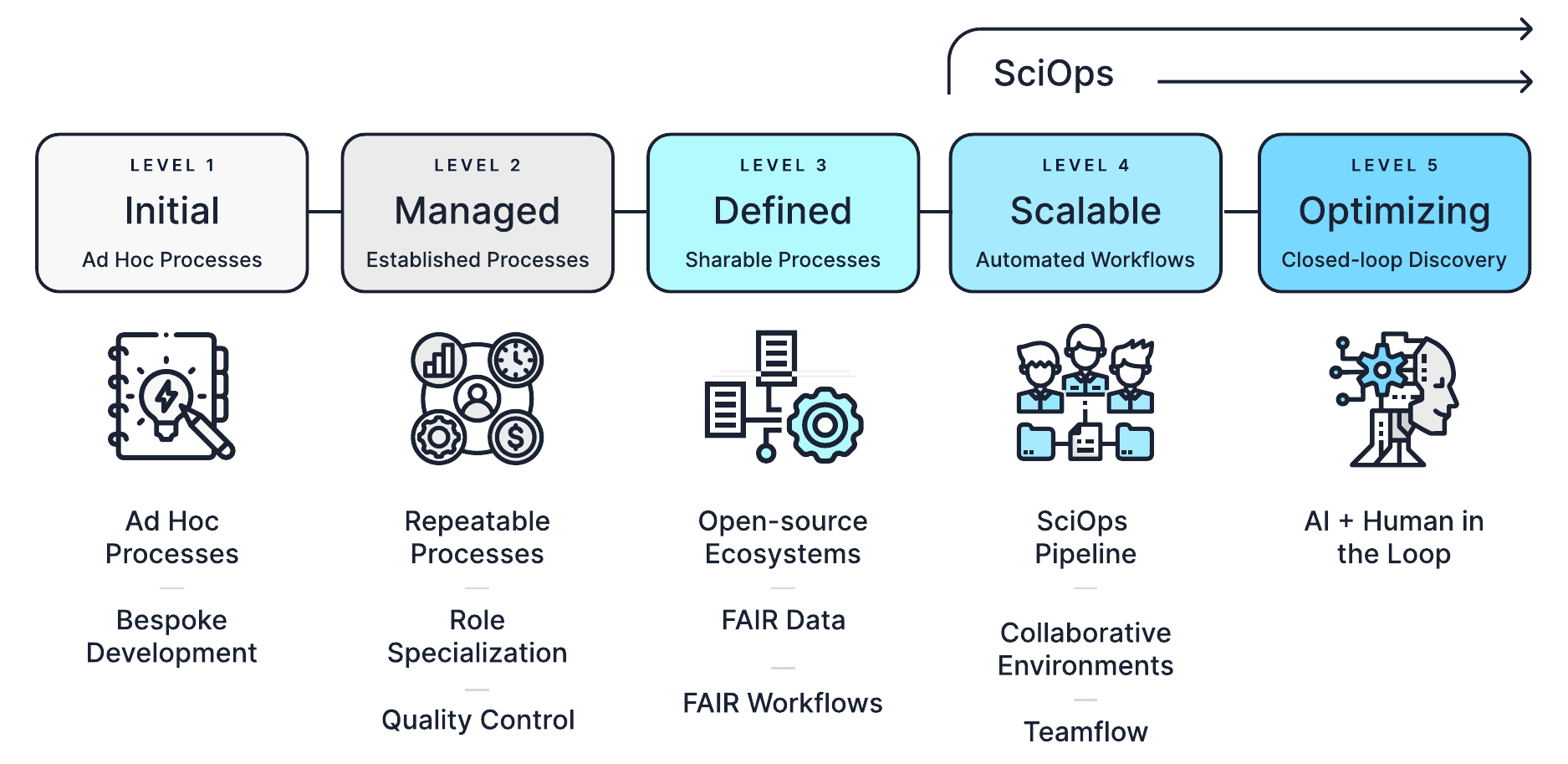}
	\caption{
The Capability Maturity Model for Science Operations.
The term ``SciOps'' describes advanced capabilities emerging at Levels 4 and 5.
} \label{fig:sciops}
\end{figure*}

Drawing on our combined experience coordinating large-scale research collaborations in neuroscience, we adapt key concepts from CMMI to the unique challenges and opportunities in contemporary neuroscience projects, without limitation to other disciplines. 
We introduce the Capability Maturity Model for Science Operations (Fig.\ \ref{fig:sciops}).

This model categorizes research teams into five levels of maturity based on their approach to planning and executing critical activities.
These levels are assessed across multiple criteria: team structure, formal processes, software implementation, data management, and computational infrastructure and procedures.
The model serves as a step-by-step guide, helping teams identify necessary steps to enhance their capabilities and scale their research efforts effectively.

Importantly, higher levels of operational maturity are not universally necessary or advantageous.
In this way, maturity levels are similar to the use of Biosafety Levels to regulate safe handling of biological hazards: each level signifies readiness for more complex projects but is not needed for less demanding scenarios.
Labs may need to advance to the next level of maturity in preparation for more demanding projects.
Furthermore, maintaining space for intuition-driven and flexible activities is crucial at all maturity levels, while successful methods and practices can be standardized and scaled up as research demands increase.

Most neuroscience teams currently operate between Levels 1 (Initial) and 2 (Managed), with some variation across subfields.
For example, operations in human neuroimaging tend to be more mature than those in experimental neurophysiology thanks to a convergence of data standards and methods \cite{bush2022lessons, kiar2023align}.
Teams at Levels 1 and 2 produce impactful findings, yet their lack of standardized processes and limited sharing can restrict their effectiveness in larger multi-lab teams and interdisciplinary initiatives.

Funding policies and publisher mandates for open data and reproducible results are driving research teams to adopt practices that advance them to Level 3 (Defined) operational maturity.
This level  emphasizes adherence to community standards and promotes reproducible, collaborative processes.

By Level 4 (Scalable), teams implement research automation, scalable computing, and efficient workflows, achieving a standard of operations described as {\bf SciOps}.
While often more achievable in larger, centralized institutions, smaller teams and individual single-PI labs can also implement SciOps practices by leveraging advanced tools and platforms.
By doing so, these smaller groups can integrate more effectively into larger research efforts, supporting broader collaboration and scalability. 

However, the pinnacle of operational maturity---Level 5 (Optimizing)---which involves closing the discovery loop with the assistance of artificial intelligence to accelerate breakthroughs, remains an aspirational goal.

The following sections detail each level, providing insights into how teams can progress at each stage.

\subsection{Level 1: Initial}
At the outset of scientific endeavors, such as establishing a new laboratory, embarking on a Ph.D. project, or initiating a novel experiment, teams typically find themselves at Level 1 of the maturity model. This stage is characterized by a high degree of flexibility and customization in experimental and analytical methods. Data volumes and throughput at Level 1 tend to be relatively small.
Each project adopts tailored approaches, with custom software and manual data management on dedicated lab infrastructure.
Standardized methods are largely absent at this stage.

\subsection{Level 2: Managed}
Advancing to Level 2 is a crucial step for research teams aiming to tackle larger, more complex projects.
At this level, the focus is on developing lab-wide standard processes that enhance consistency and predictability in internal operations.
These standards facilitate effective teamwork, make project execution more predicable, and lead to more credible findings.

To achieve Level 2, research teams must establish the following key operational characteristics: 1) establish standard protocols and repeatable processes, 2) define roles and responsibilities, and 3) establish rigorous and continuous quality controls (Table \ref{tab:level2}).

\begin{table} [h!]
	\begin{center}
		\begin{tabular}{ m{2.4cm}  m{13.2cm} }
\hline
\\
			\makecell[l]{{\bf Repeatable}\\{\bf processes}} & A level 2 team establishes uniform methods and protocols applicable across various projects. This standardization extends to data management with shared storage, standardized formats, and structured naming conventions, ensuring data integrity. Software practices also evolve, incorporating version control, documentation, testing, and code review processes \cite{artaza2016top,eglen2017toward,bush2022lessons}. A notable practice at this level is the development and maintenance of a stable data acquisition pipeline, optimized for efficient, long-term use. \\
\\
			\makecell[l]{{\bf Role}\\{\bf specialization}} & The structure of the team evolves to become more collaborative. The team defines roles and responsibilities for an efficient division of labor maximizing the use of individual expertise. A common practice is rotating trainees through different roles to provide a comprehensive understanding of lab operations.

			  The team provides a structured onboarding process for new members as well as ongoing training initiatives. These programs aim to keep the team proficient in lab operations and abreast of the latest developments.\\
\\
			\makecell[l]{{\bf Quality}\\{\bf control}} & The team establishes rigorous procedures for continuously monitoring and validating the accuracy and reliability of experimental results. These include instrument calibration, software testing, and signal quality assessment. Quality control criteria are established and periodically updated to ensure the highest standards. \\
\\
			\hline
		\end{tabular}
	\end{center}
	\caption{Characteristics of Level 2 Teams}
	\label{tab:level2}
\end{table}

The progression to Level 2 marks a significant step towards operational maturity in a given lab, characterized by a systematic approach to research, a focus on quality and reliability, and an emphasis on continuous learning and improvement.

\subsection{Level 3: Defined}
Level 3 describes research teams that embrace practices for robust collaborations across laboratories and disciplines through community standards.
Level 3 labs excel in joining forces within multi-laboratory consortia, reproducing each other's methods, streamlining interaction, and harmonizing data processes.
Key features of this level include 1) adopting open-source ecosystems for software tools and resources, 2) adhering to the FAIR principles for scientific data, and 3) establishing FAIR workflows (Table \ref{tab:level3}).

Level 3 teams, exemplified by projects such as the International Brain Lab and participants in the NIH BRAIN Initiative U19 program, demonstrate how open science practices significantly enhance operational capacity\cite{international2023modular}.
They view open science not as a burden but as an opportunity to become more efficient, credible, and accessible in their research endeavors.
This level represents a significant step towards a more integrated and collaborative scientific community, where shared knowledge and resources propel research to new heights.

\begin{table} [h!]
	\begin{center}
		\begin{tabular}{ m{2.4cm}  m{13.2cm} }
			\hline
\\
			\makecell[l]{{\bf Open-source}\\{\bf ecosystems}} & Level 3 teams are deeply engaged with resilient open-source software. These ecosystems are not just about providing tools that are ``free to use'' but also involve responsive community governance that sets standards for quality, reliability, and reproducibility. They form the foundation for state-of-the-art projects and offer community support and educational resources. Level 3 teams align their work with these community-driven open-source endeavors, promoting consistency, integration, reliability, reproducibility, and sustainability. These teams adopt disciplined practices for software management, including principled code management, peer review, validation, and testing. This approach enables the creation of evolving data pipelines and computational workflows with minimal downtime, ensuring continuous research progression. \\
\\
			\textbf{FAIR data} & Level 3 teams develop, adopt, and promote harmonized initiatives for data standards, fostering interoperability of tools and processes across research groups. This includes adherence to the FAIR principles (Findable, Accessible, Interoperable, and Reusable) for scientific data\cite{wilkinson2016fair}. Such standards facilitate the exchange and reuse of complex data, supported by robust systems including data sharing platforms that enable reproducibility and re-analysis. Examples in neuroscience include data standards created by the BIDS\cite{gorgolewski2016brain} and NWB projects\cite{rubel2022neurodata}. Data exchange and reuse requires infrastructure. Robust data sharing platforms and collaborative research environments are set up to facilitate joint research endeavors. Neuroscience data archives such as DANDI\cite{subash2023comparison}, brainlife.io\cite{hayashi2024brainlife}, BossDB\cite{hider2022brain}, and OpenNeuro\cite{markiewicz2021openneuro} not only store the data but also facilitate reproducibility and new analysis. Distributed data management systems such as DataLad, which relies on git-annex, facilitate not only versioning of data but also unified data access and exchange across multiple work sites and archives\cite{halchenko2021datalad,kalantari2023establish}. FAIR principles foster efficient work for both humans and machines. Seamless automation and effective machine learning rely on machine readability enabled by FAIR data standards\cite{huerta2023fair}. On a global scale, the collective output of teams operating at Level 3 or higher contributes to a semantic web of datasets and methods, enabling further aggregation of knowledge. \\
\\
		\makecell[l]{{\bf FAIR}\\{\bf workflows}} & Level 3 teams extend the application of FAIR principles beyond data management to encompass the entire computational workflow, tracking all data transformations from raw data acquisition through processing and analysis to the final figures in a paper \cite{goble2020fair, deelman2018future}. These practices involve managing associated code versions, dependencies, environment configurations, and parameters. Data outputs include provenance information, detailing their lineage from the original inputs through all computational transformations. This comprehensive approach ensures that computational analyses are repeatable and shareable, minimizing the potential for human error.

Formal workflows incorporate best practices for testing code logic, including unit, regression, and integration testing, often utilizing benchmark datasets. This rigorous testing framework enhances the reliability and trustworthiness of research results. The adoption of formal workflow specifications varies across scientific domains. Geosciences and bioinformatics have been at the forefront, rapidly adapting and benefiting from these advancements. The life sciences, despite inherent challenges in standardization, have also made significant progress \cite{wratten2021reproducible}.\\
\\

			\hline
		\end{tabular}
	\end{center}
	\caption{Characteristics of Level 3 Teams}
	\label{tab:level3}
\end{table}

\subsection{Level 4: Scalable}
Level 4 research operations adopt technology-enabled methods to streamline and scale collaborative efforts through semi-automated activity pipelines ({\em SciOps pipelines}) in collaborative research environments, enabling continuous project operation and efficient team workflows (Table \ref{tab:level4}).

\begin{table} [h!]
	\begin{center}
	\begin{tabular}{ m{2.4cm}  m{13.2cm} }
	\hline
\\
	\makecell[l]{{\bf SciOps}\\{\bf pipeline}} &
SciOps methodology organizes collaborative research activities into a continuous flow where team contributions are integrated through automated quality controls under collective decision making.
It unifies experimental design, data collection, processing, analysis, and dissemination into a seamless, repeatable pipeline that enhances efficiency, reproducibility, and scalability in scientific research.
A {\bf SciOps pipeline} is a technology-enabled, collaborative workflow designed to streamline scientific operations by integrating automation, data management, and continuous processes across the entire research lifecycle.
The following activities can be integrated into a shared SciOps pipeline:

\textbf{Experimental Design} defines objectives, methodologies, and resources using electronic lab notebooks and project management tools, ensuring consistency and data integrity throughout the pipeline.

\textbf{Automated Experimentation} runs experiments using automated systems for precision and scalability, with automated checks to verify experimental parameters.

\textbf{Data Collection and Aggregation} gathers and standardizes data in a shared pipeline, ensuring consistency and creating a reliable system of record for downstream analysis.

\textbf{Data Processing} automates the cleaning and transformation of raw data, maintaining integrity and supporting the continuous evolution of analysis methods.

\textbf{Data Analysis and Modeling} employs automated tools and algorithms, facilitating seamless integration of new methods and collaboration.

\textbf{Continuous Integration and Deployment for Analysis} automatically tests and integrates changes to analysis scripts and models, ensuring rapid, reliable deployment. This includes benchmarking, monitoring new data, and verifying experimental parameters.

\textbf{Collaboration and Sharing} ensures data, models, and findings are accessible to collaborators, enhancing peer review and collective learning.

\textbf{Monitoring and Feedback} provides transparency and observability of experiment performance and analysis, supporting iterative refinement and peer review. FAIR data principles preserve data lineage and provenance.

\textbf{Security and Compliance} follows legal and ethical standards for data privacy and security, employing best practices like role-based access controls.

\textbf{Infrastructure and Scalability} leverages cloud and high-performance computing for scalable data handling, with Infrastructure-as-Code enabling portability across diverse physical infrastructures.
\\
\\
	\makecell[l]{{\bf Collaboration}\\{\bf environments}}  & Level 4 teams establish streamlined collaboration environments that enable diverse, distributed teams to access data exploration capabilities.
This includes the use of web-based environments for exploratory analysis and knowledge exchange, user-friendly interfaces for data import/export, and the use of community data archives and software repositories\cite{milewicz2023devops}.  \\
\\
	\textbf{Teamflow}  & Level 4 teams establish efficient project management and communication strategies to support large-scale, multidisciplinary teams.  Project management frameworks are adopted to lower the barriers to participation. They prioritize project continuity and fair credit tracking for individual contributions. These teams foster open, transparent, and efficient communication while ensuring data consistency and integrity. Leadership and mentorship activities promote flexibility and adaptability. Integration of rapid software development practices, efficient data management techniques, and collaborative tools enable teams to work effectively and cohesively within a complex and scalable research environment.\\
\\
	\hline
	\end{tabular}
	\end{center}
	\caption{Characteristics of Level 4 Teams}
	\label{tab:level4}
\end{table}

Before SciOps, research teams experience delays and inefficiencies due to disjointed activities.
For example, data scientists would develop models and wait for software engineers to deploy these tools for experimentalists to use, requiring multiple milestones and meetings.
SciOps eliminates these issues by establishing a central activity pipeline and \emph{data pipeline}, integrating data management, code management, computational infrastructure, and collaboration tools.
This allows continuous access to primary experimental data, enabling seamless integration and direct operation by all team members.

SciOps uses improved experimental tools, computing infrastructure, and better organizational procedures to increase operational pace.
This approach enables large-scale, high-throughput experiments, iterative design, and integration of machine learning tools for data processing and analysis.
Similar to the way DevOps became widespread in diverse software teams thanks to platforms such as GitHub, adoption of SciOps practices will spread through collaborative software platforms specialized in research operations.
The SciOps formalism, automation and standardization tools developed at Level 4 enable teams to progress to new types of large-scale experiments incorporating closed-loop, computational decision making (Level 5).

To date, only a select few neuroscience research teams approach Level 4 maturity as the necessary innovations are still in development.
Among those that have made progress in this direction are well-funded institutes and multi-institution consortia, including the Allen Institute, e11 Bio, major BRAIN Initiative initiatives like BICCN \cite{brain2021multimodal} and MICrONS \cite{microns2021functional}, as well as the International Brain Laboratory \cite{abbott2017international}.
The Virtual Brain Project enables scalable and reproducible research with open-source containerized cloud services for multiscale brain simulation and magnetic resonance image processing consistent with Level 4 operations \cite{schirner2022brain}.
Platforms such as the brainlife.io\cite{hayashi2024brainlife}, the Virtual Research Environment\cite{vre} and EBRAINS\cite{ebrains}, along with the associated federated Health Data Cloud\cite{hdc} peer-to-peer networks\cite{ebrainhealth}, allow researchers to collaborate protecting data privacy or even building human digital twins for medical research.
These entities are at the forefront of neuroscience data scale and collaboration towards shared research goals and Level 4 operations.
However, significant challenges remain for the field to create an ecosystem of tools, standards, and platforms that will allow diverse research teams of all sizes and funding levels to adopt these scalable approaches.

\subsection{Level 5: Optimizing}
Level 5 represents a leap in scientific operations by bringing automation and artificial intellence into tasks that were once considered inherently human, such as making observations, generating hypotheses, designing experiments, summarizing findings, and prioritizing activities.
This level of automation enables what is described as ``closing the discovery loop,'' where processes become self-optimizing and scalable \cite{national2022automated, wang2023scientific}.

At Level 5, the goal of SciOps is to integrate artificial intelligence with human cognition, forming a dynamic partnership that refines experimental design, enhances knowledge synthesis, and supports continuous learning through computational modeling and integration (Table \ref{tab:level5}).

\begin{table} [ht]
	\begin{center}
	\begin{tabular}{ m{2.4cm} m{13.2cm} }
	\hline
\\
	\makecell[l]{{\bf Adaptable}\\{\bf experiments}}  & Level 5 teams automate experimental processes, including data collection and analysis, including experiment execution, where instruments for experiment control, stimuli, sensors, and data acquisition are integrated into the data pipeline and workflow management. \\
\\
	\makecell[l]{{\bf Machine learning}\\{\bf in the loop}} & Closed-loop experiments create a continuous automated feedback loop where machine learning algorithms optimize experiment controls to maximize knowledge gain about brain structure and function. This may involve a ``digital twin'' paradigm, where a simulation of the brain is continuously refined to match recorded data, facilitating concurrent \emph{in silico} experiments. These closed loops can span multiple temporal scales: from real-time control of experiments to long-term adaptations exploring the hypothesis space, effectively guiding the scientific study. Machine learning models powering closed-loop discovery will undergo rapid development, requiring adopting industry-proven MLOps practices whereby new models are tested and deployed continuously. Integrated AI systems observe trends in data, identify new phenomena, propose hypotheses and new experiments. Future generative AI systems will consider various information sources, including current literature, existing open datasets, and ongoing discourse between researchers, embedding this knowledge into the data pipeline and providing decision support for further inquiry\cite{wang2023scientific}. \\
\\
	\makecell[l]{{\bf Human}\\{\bf in the loop}}   & While closed-loop experiments automate many processes, human input and insight remain integral. Level 5 workflows feature a symbiotic relationship between AI tools and human ingenuity. This hybrid approach enables rapid hypothesis generation, adaptation of experimental procedures, effective quality control, and the rapid analysis of large-scale data collections. To support human participation, the data pipeline and experiment controls are made observable and explainable.  \\
\\
	\hline
	\end{tabular}
	\end{center}
	\caption{Characteristics of Level 5 Teams}
	\label{tab:level5}
\end{table}

To contrast the concept of closed-loop discovery with the conventional open-loop process, consider how the scientific process is typically described in classical training.
It is often portrayed as a linear sequence of {\em deductive reasoning}: starting with hypothesis generation, followed by experiment design to test these hypotheses, and concluding with refining and disseminating new findings.

\begin{figure*}[h!]
	\centering
	\includegraphics[width=0.5\textwidth]{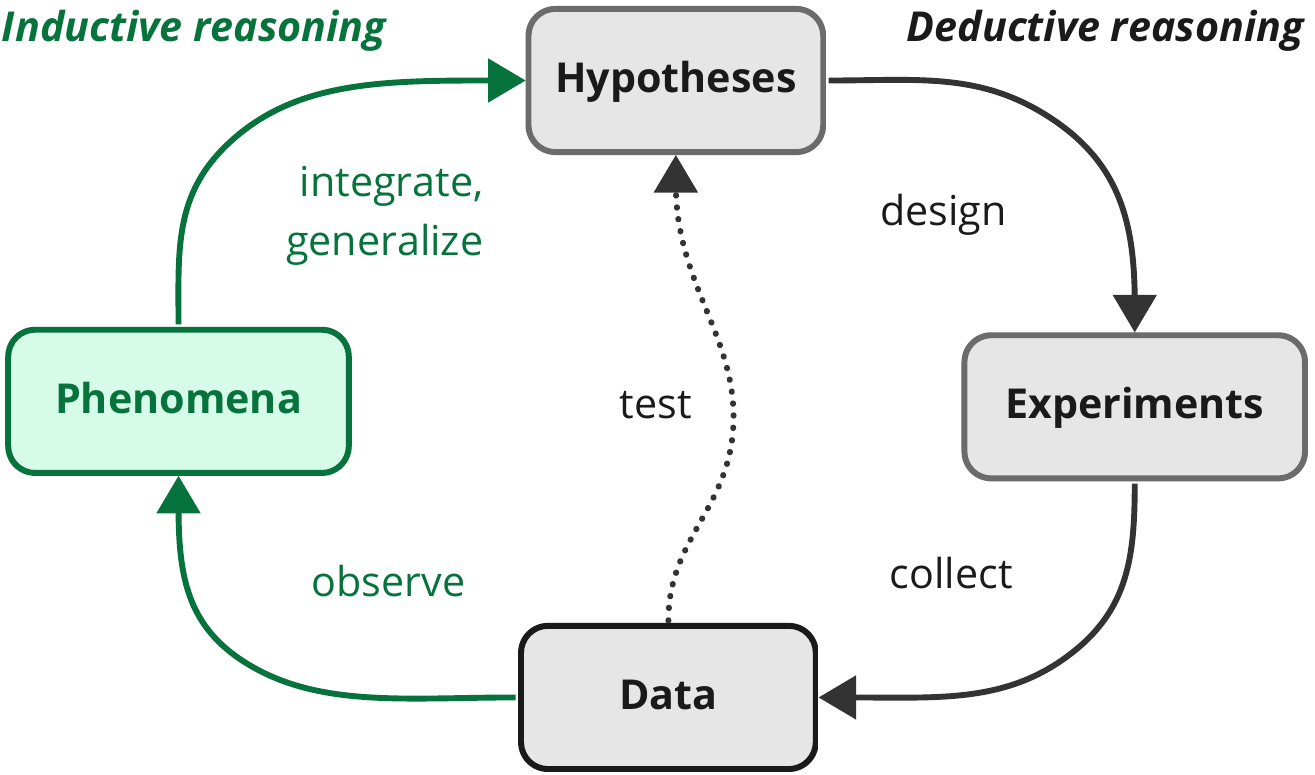}
\caption{
A schematic representation of scientific activities depicted as a ``discovery loop'' combining inductive (data-driven) and deductive (hypothesis-driven) reasoning.
Closing the loop requires automation across the entire cycle.
}
	\label{fig:loop}
\end{figure*}

However, a more comprehensive perspective views scientific discovery as an iterative cycle, often referred to as the ``discovery loop'' (Fig. \ref{fig:loop}).
This model incorporates both deductive reasoning---testing specific predictions based on general hypotheses---and inductive reasoning, which involves deriving general principles from a variety of observations.

Although this is a simplified representation of scientific activity, it serves to illustrate the concept of ``closing the discovery loop.''
The discovery process can be enhanced by applying formal methods and automation to support both the deductive and inductive phases, creating a more dynamic and continuous feedback cycle in scientific exploration.

The discovery loop is considered ``open'' when formal methods and automation are limited to the deductive part of the process, while inductive reasoning remains a uniquely human activity, lacking formal structure and automation.

Open-loop scientific studies can take the form of {\em hypothesis-driven} or {\em data-driven}:

{\bf Hypothesis-driven studies}, typically designed to investigate a specific phenomenon, testing specific predictions and models.
These are more commonly conducted by small teams of investigators typically operating at Levels 1 and 2.
An example experiment might test hypothesized mechanisms for sensory coding using an established animal model. 

{\bf Data-driven studies} are designed to create new and unique datasets such as brain atlases, for example, focusing on the data collection segment  of the deductive arc.
In neuroscience, these require higher operational maturity, Levels 3-4, to scale activities successfully to larger multidisciplinary teams.
Examples of data-driven experiments include large brain atlases and maps created by the NIH BICCN program \cite{hawrylycz2023guide} or the MICrONS program \cite{microns2021functional}. 

When dealing with complex problems using vast datasets, like reverse-engineering the brain, relying solely on human cognition to close the discovery loop becomes a bottleneck, slowing progress.
However, advances in AI are changing perceptions of its role in driving discovery and augmenting human abilities.

While classical statistics was designed for hypothesis testing (deductive reasoning), \emph{Data Science} has emerged to learn from data and uncover new patterns that support inductive reasoning.
New AI tools can help ``close the discovery loop'' by formalizing and automating inductive processes, such as synthesizing literature or adding meaningful labels to data based on established ontologies.

The rise of large generative AI models has increased confidence in AI’s ability to not only identify patterns in massive datasets but also integrate findings with existing knowledge, guide experimental studies, generate hypotheses, and support decision-making.
Human input remains crucial, in evaluating and steering the study at the high level.
For systematic exploration of well-understood experiment design, AI may quickly generate useful parameters for subsequent experiments.
In more exploratory areas, human effort will critical in interpreting new data and relating it to literature to decide on next steps with AI assistance.

This shift marks a major change in scientific methodology, moving from human-driven data exploration to a collaborative model where AI and human researchers work together to accelerate discovery \cite{gao2024empowering}.

While no team has fully achieved Level 5 operations, some projects that optimize experimental conditions using machine learning already demonstrate aspects of this vision.
Small-scale closed-loop experiments have long been used to study dynamic brain phenomena, where the experiment is adjusted based on real-time analysis. The feedback period can range from near real-time to several minutes or even days.
In these experiments, sensory stimuli, experiment parameters, and brain activity are actively controlled through feedback from ongoing analysis of behavior or neural activity \cite{tumer2007performance, dimattina2013adaptive}.
More advanced experiments have even used detailed neural network models to create ``digital twins'' of the brain.
These models allow for accelerated experiments and inferences from recorded data, guiding the direction of the experiment or even the entire study \cite{walker2019inception}.

Future closed-loop studies will use highly configurable experiment designs and automated machine learning tools to discover new patterns in the data and guide experiments for optimal knowledge gain \cite{greenhill2020bayesian, paninski2018neural, lorenz2016automatic}.
Human researchers will play a key role in generating hypotheses, designing experiments, setting success metrics, and steering priorities to ensure rapid validation and evaluation of results.
Longer AI-driven feedback loops will also incorporate broader contexts, such as the latest research findings, ongoing collaboration, and alignment of results across multiple experiments.

\section{Research Communication and Publishing}

The efficacy of scientific research depends greatly on how effectively it is evaluated, communicated, and disseminated.
{\bf Research communication\*} is a critical component of scientific operations, serving not only to share new discoveries but also to build professional reputations, track merit for career advancement, and inform hiring and funding decisions.
As research operations become more dynamic and complex, communication practices must evolve in parallel to meet these growing demands.

The traditional publishing model, centered on scientific journals, has long been instrumental in maintaining research quality through peer review, assigning credit, and providing metrics to assess contributions.
However, the model is increasingly facing limitations that hinder the effective acquisition, validation, and communication of new knowledge.
Rooted in a print-based format, it struggles to accommodate the data-rich and interactive nature of modern research, making it challenging to convey complex and evolving findings.
While some enhancements---such as supplementing journal articles or preprints with data files or links to data and code repositories---have been introduced, these efforts remain constrained by the legacy of print-based workflows, placing significant burdens on researchers.

Furthermore, the traditional publishing model often falls short in recognizing the diverse contributions involved within large-scale, multidisciplinary projects and offers little incentive for ensuring the reproducibility of methods, software, or techniques outside the originating lab.
Negative results, despite their scientific importance, remain underreported in published literature, leading to gaps in data availability for secondary analysis \cite{fanelli2012negative}.
The slow pace of the publishing process, combined with the fear of ``getting scooped'' and delayed or missed credit assignment, discourages the timely exchange of ongoing research and inhibits real-time collaboration.

Additionally, the print-based format limits accessibility for meta-analysis or AI-driven synthesis, restricting the integration of knowledge across disciplines.
As research operations advanced toward higher levels of operational maturity, the traditional publishing model struggles to keep paces, highlighting the need for more modern, integrated, and scalable communication practices.

New alternatives are emerging. 
{\bf Preprint servers} allow researchers to share findings before formal peer review, accelerating dissemination and encouraging early feedback.
{\bf Open-access publishing} broadens the reach and impact of research while promoting transparency.
Digital repositories enable the sharing of datasets, code, and methodologies in standardized formats, facilitating meta-analysis, AI-driven processing, and reproducibility.
These innovations reflect a shift towards real-time data sharing as an integral part of research operations, where metadata on collaboration and peer assessment can be mined by AI to provide ongoing snapshots of the research process.

More innovative approaches are emerging, however, with dynamic publications that integrate living, data-driven figures and embedded code, bridging the gap between research execution and communication. Notable examples include the American Geophysics Union’s {\em Notebooks Now!} project \cite{caprarelli2023notebooks} and reproducible preprint platforms like NeuroLibre \cite{karakuzu2022neurolibre}. These initiatives allow research to be shared more quickly, with projects like Aligning Science Across Parkinson’s (ASAP) reporting an average reduction of 5.5 months in time to publication by using preprints \cite{parkinsonsresearchpublishing}.

In response to the evolving research landscape, new models for credit and recognition are also being explored. The National Academies Roundtable on Aligning Incentives for Open Scholarship is actively promoting changes in academic and funding structures to incentivize open science \cite{national2024promoting}. Additionally, tools like PLOS Open Science Indicators \cite{plosindicator} and the European Union’s Open Science Monitor \cite{eumonitor} track and promote global trends in open access, collaboration, and transparency. While these open-science practices have been shown to accelerate research, traditional metrics for scientific contribution still undervalue open and collaborative work. As new incentives and infrastructure emerge, research communication will continue to evolve, challenging the current publishing model and fostering a more open, efficient, and collaborative research environment.

There is a clear opportunity to develop tools that better integrate research communication into the research workflow itself, streamlining the process and enhancing transparency, collaboration, and efficiency. These tools could help bridge the gap between research execution and communication, facilitating real-time data sharing, reuse, and reproducibility in an increasingly digital and data-driven research environment.
SciOps platforms supporting Level 4 and 5 operations, must be built with research communication in mind. 

\section{Advancing Modern Scientific Operations}

The evolution of scientific practices, coupled with rapid technological advancements, demands a strategic re-organization in how we approach research, especially in the context of experimental, big-data neuroscience. The transformational power of SciOps methodologies has the potential to reshape the way we think about scientific endeavors through closed-loop, scalable experiments which provide new insight into neurological disease, neural information processing, and more. However, to tap into this potential, the community must focus on specific areas of development:

\subsection{Action 1: Adopt the Capability Maturity Model for Scientific Operations}

\textbf{Community Governance:} We invite the community to embrace, enhance, and collaboratively govern this Capability Maturity Model.
This first version of the Capability Maturity Model for Scientific Operations, which we can designate as version 1.0, serves as a starting point, and we are committed to establishing a roadmap and guidelines for contributions, working in coordination with the International Neuroinformatics Coordinating Facility (INCF)\cite{abrams2022standards}.

A centralized community resource will be established through the INCF and other platforms to provide access to the maturity model and related resources, including a community managed website with contribution guidelines. Documentation about and reference examples of the model will serve as a valuable reference point for individuals and organizations interested in advancing scientific operations. To facilitate community engagement, we envision the creation of a dedicated SciOps resource and a working group to administer and support the framework and its collaborative development, including the creation of future versions of the Capability Maturity Model.

\textbf{Assessment and Roadmaps:} The model will support assessment and certification processes, enabling organizations to prepare their processes for various projects and programs. It will also serve as a valuable tool for charting organizational improvements and technological roadmaps.

By embracing this Capability Maturity Model and actively participating in its development and application, the scientific community can collectively advance the field of scientific operations, foster innovation, and drive transformative progress in research methodologies and practices.

\textbf{Communicate expectations across the neuroscience community:} For this model to be adopted, the expectations around how neuroscience needs to be performed to meet 21st century goals will have to be socialized across the neuroscience community, through integration with training programs, townhall meetings and workshops at major scientific gatherings.

\subsection{Action 2: Establish SciOps Methodologies}

Current standards and policies in neuroscience data focus on standardization and public data sharing, marking progress towards Level 3 maturity.
Achieving Levels 4 and 5 requires new tools and ``SciOps methodologies,'' which adapt DevOps principles to neuroscience experiments.
Many practices can be transferred from collaborative scientific workflow management systems prominent in bioinformatics \cite{wratten2021reproducible, di2017nextflow, koster2012snakemake, vivian2017toil} and from industry DevOps and DataOps frameworks and platforms \cite{magicquadrant}.

\textbf{Experiment Automation:}
Workflow management technologies should seamlessly integrate with neuroinformatics tools and methods, setting the stage for scalable operations as defined in Levels 4 and 5.
The evolution of neuroinformatics tools should prioritize continuous integration and deployment of research software—whether commercial or community-driven—encompassing experiment design, data acquisition, and analysis.
Data should be available in formats and on infrastructure that allow for scalable storage, processing, and sharing.
As data scales, formats and infrastructure evolve rapidly, making it critical to establish lasting organizational principles for project continuity.
Efforts may be required to focus on integrating existing data standards with emerging AI tools to provide the ontologies and data specifications required for AI-driven neuroscience.
Additionally, experimental workflows should integrate formal frameworks for embedding artificial intelligence into the discovery loop.

Software tools tailored specifically for SciOps can simplify intricate processes, streamline tasks, and significantly enhance research efficiency.
With investment in new software tools and automation approaches, we can usher in a transformative era of scientific operations characterized by heightened methodological prowess and overall efficacy.

\subsection{Action 3: Focus on Digital Platforms}

Digital platforms have permeated all spheres of life, providing virtual spaces for efficient interactions under uniform organizational models.
`Platformatization' has come to dominate many aspects of academic research, including archives, repositories, scientific gateways, publishing, data collection for citizen science, and academic social networks \cite{dasilvaneto2023platform}.
Large-scale grid projects across scientific domains have established non-profit organizational models, including customer service, support, and training (SBGrid \cite{morin2013collaboration}, XSEDE \cite{towns2014xsede}).
However, internal operations in active phases of scientific projects still rely on home-made processes in both academic and commercial research, limiting their operational maturity.

Digital platforms can elevate scientific operations without requiring extensive investments in engineering expertise and custom solutions, democratizing complex scientific operations.
The neuroscience community has defined recommendations for next-generation platforms, or scientific gateways, which integrate services while promoting transparency and accessibility \cite{sandstrom2022recommendations}.
A new generation of academic neuroinformatics platforms, such as {\tt brainlife.io} \cite{hayashi2024brainlife}, DABI \cite{duncan2023data}, OpenNeuro \cite{markiewicz2021openneuro}, SPARC \cite{bandrowski2021sparc}, and DANDI \cite{rubel2022neurodata}, are poised to replicate the successes seen in bioinformatics and biomedical research (e.g., Galaxy \cite{afgan2022galaxy}) and structural biology (e.g., CCP4 \cite{krissinel2022ccp4}).
To achieve this, they must evolve beyond data sharing (Level 3) to the automation and scalability of workflows (Levels 4 and 5).

Academic digital platforms often suffer from limited usability, creating the perception that shared infrastructure can impede daily activities and hindering sustained adoption.
Even when data standards and shared infrastructure projects are available, a lack of training and programming skills limits the uniform adoption of tools and data standardization \cite{schottdorf2024data}.
As a result, some teams build home-made solutions rather than adopting centralized platforms endorsed by their communities.
Continuous improvement is required in service quality, performance, customer support, and long-term sustainability.

Commercial platforms, driven by competitive pressures, strive to deliver usability, robust customer support, and service continuity.
In the Life Sciences industry, a new generation of commercial software platforms is emerging to enhance scalability and reproducibility in large-scale research operations \cite{effron2022changingworld, deeporigin2024softwarelandscape}.
These platforms aim to improve scientific workflows (such as electronic laboratory notebooks and laboratory management tools) and data infrastructure for storage, management, and harmonization.
Increasingly, they enable the use of AI for scientific discovery, where machine learning tools built on modern computing infrastructure accelerate neuroscience analysis and experimental design.
In neuroscience, emerging commercial platforms such as DataJoint Works, Inscopix IDEAS, and CodeOcean provide support for integrated research workflows, including experimental data management and reproducible computational workflows.

To ensure the success of these platforms, it is essential that they integrate seamlessly with FAIR data and FAIR workflows, emphasizing transparency, accessibility, and interoperability of data and computational services.
The maturity model can assist commercial and academic technology developers in aligning their tools and platforms with the overarching objective of expanding the research capabilities of neuroscience research teams.
Given the diverse nature of the field, it is unlikely that a single platform or toolset will dominate.
Diverse groups with common goals and roadmaps can form alliances around shared standards and open-source frameworks, promoting interoperability, transparency, and reproducibility across their respective platforms.

To realize this vision, it is essential to establish a comprehensive strategy that encompasses a blend of academic projects, commercial technology initiatives, and consortial activities, all aligned with funding policies.
This multifaceted approach will guide the formulation of research projects, the marketing efforts of commercial technology providers, and the policies of funding agencies, ultimately fostering the creation of a unified and sustainable ecosystem.

\section*{Acknowledgments}
We would like to thank the many community members who contributed to the discussions and reviews of this manuscript and raised key community issues, including David Feng, Andreas S. Tolias, Forrest Coleman, Marisel Villafañe-Delgado, Lindsey Kitchell, Daniel Xenes, Mathew Abrams, Taige Abe, Dan Birman, Srinivas Gorur-Shandilya, Ryan Kosai, Nabil Laoudji, and many others.
DY, BW, TN, MK, KG, and ECJ were supported in part by the NIH (award R44 NS129492). YOH and SG were supported in part by the NIH (awards 1 R24 MH117295 and 2 P41 EB019936-06A1 R). The content is solely the responsibility of the authors and does not necessarily represent the official views of the National Institutes of Health or other supporting institutions.

\section*{Author Contribution Statements}
All authors contributed to the conceptualization, writing, and editing of the manuscript. ECJ, FP, and DY also were responsible for final aggregation and editing of the manuscript, as well as figure creation. All authors agree with the manuscript content. 

\section*{Competing Interests}
DY, MK, TTN, and KG have equity interest in DataJoint Inc.

\section*{Data and Code Availability}
No new data or code were developed for this work, and availability of individual software tools and datasets are described in their relevant references. 

\bibliographystyle{naturemag}
\bibliography{main}

\end{document}